# Superconductivity in a Magnetic Rashba Semimetal EuAuBi

Hidefumi Takahashi[1,2*], Kazuto Akiba[3], Masayuki Takahashi[1], Alex H. Mayo[1,4], Masayuki Ochi[5], Tatsuo C. Kobayashi[3] and Shintaro Ishiwata[1,2†]

[1]*Division of Materials Physics and Center for Spintronics Research Network (CSRN), Graduate School of Engineering Science, Osaka University, Osaka 560-8531, Japan*
[2]*Spintronics Research Network Division, Institute for Open and Transdisciplinary Research Initiatives, Osaka University, Yamadaoka 2-1, Suita, Osaka, 565-0871, Japan*
[3]*Graduate School of Natural Science and Technology, Okayama University, Okayama 700-8530, Japan*
[4]*Institute for Materials Research, Tohoku University, Sendai 980-8577, Japan.*
[5]*Department of Physics, Osaka University, Machikaneyama-cho, Toyonaka, Osaka 560-0043, Japan*

We report the observation of superconductivity with multiple magnetic ordering and Rashba-type spin-orbit coupling in a layered polar semimetal EuAuBi. Magnetic transition is observed at 4 K, followed by a superconducting transition at 2.2 K, which is sensitive to the crystal surface conditions. The upper critical field $H_{c2}$ of 9.8 T for the out-of-plane field is three times higher than that for the in-plane field, which can be associated with the two-dimensional structure or the surface state. On the basis of first-principles calculations, it is found that the characteristic $H_{c2}$ possibly reflects the anisotropic modification of the Fermi surface by the effective combination of Rashba-type spin splitting and Zeeman spin splitting enhanced by Eu moments.

The role of magnetism in superconductivity has been extensively discussed in terms of the exotic pairing mechanism[1]. This is highlighted by the high-temperature superconductivity near the antiferromagnetic quantum critical point in copper oxides and iron-based pnictides[2–5], as well as the possible spin-triplet superconductivity in the vicinity of the ferromagnetic quantum critical point[6–8]. Another factor yielding non-trivial superconductivity is the absence of spatial inversion symmetry, by which a mixture of singlet and triplet superconducting pairing is expected[9–11]. Recently, these exotic superconductors have received considerable attention owing to their topological properties[10,12]. A Majorana edge state is expected to be applied to the next generation of innovative quantum computing[13].

The questions that naturally arise in this regard are as follows: (i) how are the superconducting state and topology affected when magnetism (time-reversal symmetry breaking) and spatial inversion symmetry breaking coincide?; (ii) can the superconducting properties be controlled using external fields? In this regard, heavy-fermion noncentrosymmetric (NCS) superconductors such as Ce-based compounds can be considered[9,10]. However, the relationship between spatiotemporal symmetry breaking and topology in these superconductors is elusive because of the low superconducting transition temperature and complex Fermi surfaces, which reflect the coupling between localized *f*-electrons and itinerant electrons. Moreover, this coupling typically reduces the spin moment, thereby rendering it difficult to control the electronic state using an external magnetic field. Recent theoretical studies predicted that topological Fulde–Ferrell–Larkin–Ovchinnikov (FFLO) superconductivity can be realized in systems with broken time-reversal or space-inversion symmetry, as exemplified by doped Weyl semimetals and Rashba semiconductors[14–16]. In addition, surface superconductivity is reported in such topological semimetals, which is also influenced by the internal magnetic field.[17] Therefore, the ideal system for magnetic topological superconductors is a magnetic semimetal or a semiconductor with a polar structure, where the localized spin moments enable the manipulation of superconducting properties.

Here, we focused on EuAuBi with a LiGaGe-type structure [see Fig. 1(a)], which is one of the hexagonal *ABC*-type ternary compounds[18]. The Eu and Au/Bi atoms in EuAuBi form a two-dimensional triangular lattice and buckled honeycomb lattice, respectively. Because of this buckled structure, LiGaGe-type compounds are polar in the *c*-axis direction and have been theoretically proposed as candidates for Rashba semiconductors and Dirac/Weyl semimetals.[19,20] It is noteworthy that some of the hexagonal *ABC* compounds without the buckled structure (AlB$_2$ derivative structure) possess superconductivity.[21,22] EuAuBi contains magnetic ions of Eu$^{2+}$ ($S = 7/2$) and is expected to exhibit the magnetic order at low temperatures, as observed in the similar Eu-based *ABC* compounds such as EuCuSb[23]. In this study, we obtained single crystals of EuAuBi and discovered its superconductivity at 2.2 K, i.e., below the magnetic ordering temperature of 4.0 K. Furthermore, whereas the upper critical field $H_{c2}$ for the in-plane magnetic field (*H* // *ab*) at the lowest temperature (0.1 K) was approximately 3 T, the maximum $H_{c2}$ for the out-of-plane magnetic field (*H* // *c*) was 9.8 T. This characteristic feature of $H_{c2}$ will be discussed as an indicator of unconventional superconductivity in a magnetic polar semimetal EuAuBi with significant Rashba-type spin splitting and enhanced Zeeman splitting.

As shown in the temperature dependence of the magnetic susceptibility *M*/*H* measured with $H = 0.1$ T [Fig. 1(b)], the compound underwent an antiferromagnetic (AFM) transition at 4 K ($T_N$), which was also detected by the measurement of specific heat *C* [Fig. 1(d)]. As shown in the insets in Fig. 1(b), *M*/*H* measured with an in-plane magnetic field (*H*//*ab*) increases significantly below $T_N$, whereas *M*/*H* with an out-of-plane field (*H*//*c*) are nearly saturated and shows a weak anomaly at $T_N$, implying that the spin structure is anisotropic.[23] The AFM nature was also confirmed by the Curie–Weiss fits to the (*H*/*M*)–*T* curves for *H*//*c* and *H*//*ab* from 300 to 200 K, as shown in Fig. S1 [24], which resulted in Weiss temperatures of $\theta_w$ = -16.9 and -10.1 K, respectively. In addition to the AFM transition, the magnetic susceptibility below $T_N$ indicated weak anomalies (denoted by blank triangles in insets of Fig. 1(b), suggesting subtle spin reorientation at low temperatures. Figure 1(c) shows the temperature dependence of the in-plane (*I*//*ab*) electrical resistivity ρ. Metallic behaviour was observed with a kink at $T_N$. At 2.2 K ($T_C$), where the out-of-plane (*H*//*c*) and in-plane (*H*//*ab*) magnetic susceptibility (4πχ) measured with various fields exhibited a Meissner effect [Fig. 1(e)], ρ decreased abruptly to zero, which is characteristic of superconductivity (onset of the superconducting transition temperature $T_c$(onset) is 2.2 K). The volume fraction of superconductivity at 1.8 K is estimated to be about 80 % at lowest fields from the field cooling susceptibility (we have measured the magnetic and transport properties for another sample shown in Ref. 24 [Fig. S2 and Fig. S3]). The Meissner transition disappears by the application of very low magnetic fields of 5 Oe, suggesting that the lower critical field $H_{c1}$ is extremely small as in the noncentrosymmetric superconductor YPtBi[25]. The critical temperature $T_c$ decreased upon the application of *H*//*c*, which supports the emergence of a superconducting phase [24]. However, a clear jump corresponding to the superconducting transition is absent in the specific heat data, which reflects the small electronic specific heat coefficient inherent to the semimetallic band structure forming the pseudogap and the presence of the large magnetic entropy. On the basis of the band calculations, the density of states DOS($E_F$) near the Fermi level is estimated to be about 0.2 (PM) ~ 0.5 (FM and AFM) states/eV(see Fig. S5 in Supplemental Material), and the electronic specific heat coefficient γ is evaluated to be $(1/3) \times (\pi^2 k_B^2) \times$ DOS($E_F$) = 0.00047~0.00117 (J/molK$^2$). Therefore, the estimated jump in specific heat capacity upon the superconducting transition ($\Delta C=1.43\gamma T_c$=0.0017~0.0042 J/molK) is not observable in the presence of the fairly large magnetic specific heat capacity (~ 10 J/molK).

To clarify the magnetic properties, we measured the magnetic-field dependence of the magnetization *M* at various temperatures. Figures 2(a) and 2(b) show the magnetization measured in the *H*//*c* and *H*//*ab* fields, respectively. For *H*//*c*, the magnetization curve is almost linear and shows a smeared kink associated with magnetic saturation at approximately 6 T (= $H^c_{N1}$) at 1.8 K [Fig. 2(a)]. On the other hand, the magnetization curve for $H//ab$ shows small kinks at approximately 2 T (= $H^{ab}_{N1}$) and 3 T (= $H^{ab}_{N2}$) along with hysteretic behavior below 2 K [Fig. 2(b)], which implies spin-flop transitions. Because the in-plane saturation field of 5.3 T (= $H^{ab}_{N3}$) is slightly lower than the out-of-plane saturation field of 6 T, the localized moment is likely to behave as a Heisenberg-like spin with easy plane-type anisotropy. Since the magnetization for both field directions are not completely saturated even at 7 T (~6 $\mu_B$), it is difficult to discuss the deviation of the valence of Eu from +2 by comparing the saturation magnetization with that expected for Eu$^{2+}$ (7 $\mu_B$). However, the effective magnetic moments estimated from a Curie-Weiss fitting (7.94 $\mu_B$ for $H//c$ and 7.60 $\mu_B$ for $H//ab$) and the magnetic entropy estimated from the magnetic specific heat (17.7 J/mol K) are fairly close to the theoretical values of 7.94 $\mu_B$ and 17.3 J/mol K for Eu$^{2+}$, respectively (Figs. S1 and S4 in Supplemental Material).

The magnetoresistance was measured using magnetic fields $H//c$ and $H//ab$ [see Figs. 2(c) and 2(d)]. Below 2 K, $\rho$ increases rapidly with the magnetic fields in the low-field region (< 2 T), reflecting the suppression of the superconducting state. Upon the application of $H$ along $c$ at 3 K, weakly positive and negative magnetoresistance was observed, presumably due to the coupling between localized moments and itinerant electrons, indicating the enhancement and suppression of spin scattering in AFM1 phase and paramagnetic phase, respectively[26].

To gain more insight into the superconducting state, we conducted measurements of electrical resistivity $\rho$ at low temperatures down to 0.1 K. Inset of Fig. 3(a) shows the temperature dependence of $\rho$ below 4 K. Whereas the onset superconducting transition around 2.2 K was confirmed for several samples, sample No. 1 exhibited an unexpected increase in $\rho(T)$ below 1.8 K (denoted by $T_{unknown}$), followed by a gradual decrease in $\rho(T)$ toward the lowest temperatures (In addition to sample dependence, samples also depend on treatment, discussed in Ref. 24 [Note 6]). This reentrant-like behavior has been observed in several superconductors and discussed in terms of the microscopic Josephson junctions formed by intergrowth in single-crystalline samples[27–29] (see Fig. S10 in Ref. 24). Another potential mechanism for the reentrant-like behavior is the formation of superconducting paths due to inhomogeneities in the superconducting state possibly caused by the polar structural domains. The magnetic field dependences of $\rho$ with $H//c$ and $H//ab$ are shown in Figs. 3(a) and 3(b), respectively. $\rho(H)$ exhibited a convex-like unusual enhancement at low magnetic fields (0.1-0.2 T); subsequently, the superconducting state was recovered by the magnetic fields of several teslas, and the superconducting state returned to the normal state at ~ 10 T and 3 T for $H//c$ and $H//ab$, respectively (The complex behavior in $\rho(H)$ is typically seen for microscopic Josephson junctions). The observation of such high and anisotropic $H_{c2}$ values precludes the possibility of superconductivity originating from impurities; this is because such high $H_{c2}$ values have not been reported in alloys and nanoparticles of Bi or AuBi[30–32].

By summarizing the magnetization and resistivity data, we established the magnetic and superconducting phase diagrams of EuAuBi as functions of $H$ and $T$ for $H//c$ and $H//ab$ as shown in Figs. 4(a) and 4(b), respectively ($T_c$ and $H_c$ are determined by the onset temperatures and fields as shown in Fig. S11 in Supplemental Material). For $H//c$, the AFM and superconducting phases coexist at the lowest temperature up to 6 T. The superconducting state persists even in the induced ferromagnetic state at the lowest temperature above 6 T, where the magnetic moments are aligned nearly ferromagnetically. Meanwhile, for $H//ab$, the superconducting phase was embedded in two types of magnetic phases AFM1 and AFM2. The emergence of rich magnetic phases (AFM1-3) under a magnetic field with $H//ab$ is analogous to the related Eu-based compounds such as EuCuSb showing competing magnetic phases including a helimagnetic phase, suggesting that the magnetic ground state of EuAuBi shares a similar magnetic phase with unique spin textures.[23,33,34]. The phase diagrams suggest that the superconductivity is less sensitive to the variation of magnetic ordering, reflecting the consisting of superconducting Au-Bi layers sandwiching the magnetic Eu ions. However, the large magnetic moment of Eu ions potentially enriches the superconducting state by acting as effective internal magnetic fields, as exemplified by the $H_{c2}$ anisotropy. Furthermore, the NCS structure is expected to introduce a helical spin texture at the Fermi surface through the band splitting due to the Rashba spin–orbit coupling, which also results in unconventional magnetic responses in the superconducting phases. Prior to discussing these possibilities, we estimated the orbital limiting field $H_{orb}$ = 4.32 T using the Werthamer-Helfand-Hohenberg formula in the clean limit $H_{orb} = 0.72T_c \, [-dH_{c2}/dT]_{Tc}$, and the Pauli limiting field $H_P$ = 3.5 T using $H_P = \Delta/\sqrt{2}\mu_B$ and $\Delta = 1.76k_BT_c$ ($T_c$ = 2.0 K).[35] $H_{c2}$ ~ 10 T for $H//c$ surpasses both $H_p$ and $H_{orb}$, as observed in heavy-fermion NCS superconductors[9,36].

The anisotropy of $H_{c2}$ is typically found in quasi-two-dimensional superconducting systems, implying the anisotropic band structure. For a simple model, $H_{c2}$ is associated with the electron effective mass $m^*$ as $H^c_{c2}/H^{ab}_{c2}=(m^*_{ab}/m^*_c)^{0.5}$, where $m^*_{ab}$ and $m^*_c$ are the effective masses normal to and along the $c$-axis, respectively[37]. In quasi-two-dimensional systems, $H^{ab}_{c2}$ (for $H//ab$) is typically higher than $H^c_{c2}$ (for $H//c$) because $m^*_{ab}$ tends to be smaller than $m^*_c$. In the present system, $m^*_{ab}$ = 0.14-0.88$m_0$ and $m^*_c$ = 1.09-1.46$m_0$ are evaluated from the band calculations shown in Fig. 5(a) (see Table S1 and Fig. S4 in Ref. 24). Despite the quasi-two-dimensional structure of EuAuBi, $H^c_{c2}$ ~ 10 T is almost three times higher than $H^{ab}_{c2}$ ~ 3 T, similar to heavy-fermion NCS superconductors, which demonstrates intriguing anisotropy presumably arising from Rashba-type spin splitting[36].

To investigate the effect of the Rashba-type spin-orbit coupling and Zeeman interaction on the superconducting properties, we performed first-principles band calculations including spin–orbit couplings. Figure 5(a) shows the band structure in the ferromagnetic (FM) and antiferromagnetic (AFM) states with the Eu magnetic moments aligned along the $c$-axis (for the AFM states with the magnetic moments aligned along the $ab$ plane, see Fig. S4 in Ref. 24). The semimetallic band structure can be confirmed by the presence of hole pockets around Γ and electron pockets at M and A points. Figure 5(b) shows the Fermi surface with spin–orbit couplings in the FM phase. The cylindrical and ellipsoidal Fermi surfaces extending in the $k_z$ direction at Γ, M and A points exhibit a quasi-two-dimensional band structure, as expected for the layered structure. Therefore, the effective mass model, which yields an $H^c_{c2}$ value that is lower than $H^{ab}_{c2}$, is not suitable for explaining the characteristic anisotropy of $H_{c2}$ in EuAuBi. The origin of the unique $H_{c2}$ anisotropy similar to CeRhSi$_3$ can be reasonably explained as follows. Materials with a polar structure have a finite Rashba-type spin-splitting in the band structure through the spin-orbit coupling; in EuAuBi, the Fermi surface is formed by Bi-6$p$ orbitals with strong spin-orbit coupling [24]. In this case, the in-plane magnetic field introduces asymmetry in the in-plane Fermi surface with spin splitting, where extra momentum is involved in forming the Cooper pair, giving rise to the enhancement of the depairing effect. On the other hand, this effect should be negligible for the out-of-plane field causing the symmetric Zeeman-type spin splitting, which has nothing to do with the symmetry of the in-plane Fermi surface (see Fig. S12 in Ref. 24)[38–41]. This explanation is applicable to EuAuBi as justified by the in-plane band structures (cross section of $k_Z$ = 0) with Eu moments aligned along the $c$-axis ($m_{Eu}//z$) and $ab$ plane ($m_{Eu}//x$), as shown in Figs. S13(a) and S13(b), respectively[24]. For $m_{Eu}//x$, the Fermi surfaces at Γ and M points exhibited asymmetric deformation along the $y$-direction (-M-Γ-M) as expected from the Rashba spin-orbit system; for instance, the Fermi surfaces for $m_{Eu}//z$ and $m_{Eu}//x$ indicated by white lines differed significantly from each other. This field-dependent difference was clearly found in the calculated band dispersions with spin polarization as shown in Figs. 5(c) and 5(d). Whereas the band dispersion was symmetric for $m_{Eu}//z$ [Fig. 5(c)], the band dispersion along the -M-Γ-M direction was highly asymmetric for $m_{Eu}//x$ [Fig. 5(d)].

Finally, we discuss the possibility of nontrivial superconducting states in EuAuBi. Owing to the coexistence of superconductivity and magnetism with NCS structure, the possibility of exotic superconducting states such as triplet and mixed singlet-triplet pairing can be considered, as proposed for heavy-fermion superconductors. Based on band calculations, the characteristic feature of $H_{c2}$ is attributable to the significant spin splitting in terms of the Rashba spin-orbit coupling, which is one of the necessary conditions to achieve nontrivial superconducting states. It is noteworthy that the Rashba spin-orbit coupling and in-plane magnetic field (perpendicular to the polarity) may induce FFLO superconductivity by the asymmetric deformation of the Fermi surface[14,15]. Furthermore, such intriguing superconductivity has recently been discussed in terms of topological states such as the Majorana edge state.

Another intriguing phenomenon is the superconducting state emerging on the crystalline surface or at the polar domain boundaries [17,32,42]. The anomalous enhancement in $\rho$ below 1.8 K, which is dependent on the sample, may be relevant to the surface state and domain state ($T_c$ is also sensitive to the sample state as discussed in Ref. 24 [Note 6]). Furthermore, in the Meissner effect, although a reasonably large superconducting volume fraction of ~80% is observed at 1.8 K, this signal disappears with the application of the extremely small fields. This suggests that magnetic flux can easily penetrate even in weak magnetic fields. One possibility is the emergence of a spontaneous flux state by internal magnetization like ferromagnetic superconductors[43,44]. Another potential mechanism for the reentrant-like behavior is the formation of unusual superconducting paths due to inhomogeneities in the superconducting state, which is presumably associated with the polar structural domains. The fact that small finite resistivity remains even below the superconducting transition under high magnetic fields (> 4T) as shown in Fig. 3(a) also implies that fluctuations due to magnetic flux penetration within the crystal[45] and an unusual state such as surface superconductivity may be realized. On the other hand, surface and polar domains essentially break the inversion symmetry and can yield Rashba-like electronic states, which is consistent with the magnetic field anisotropy obtained in the experimental results. To distinguish between bulk and surface superconductivity, more detailed experiments are indispensable.

**Acknowledgements**
The authors thank S. Fujimoto, T. Mizushima, and K. Aoyama for fruitful discussions. This study was supported in part by KAKENHI (Grant No. 19K14660, 20K03802, 21H01042, 21H01030, 22H00343), the Asahi Glass Foundation, the Murata Foundation, and Research Foundation for the Electrotechnology of Chubu.

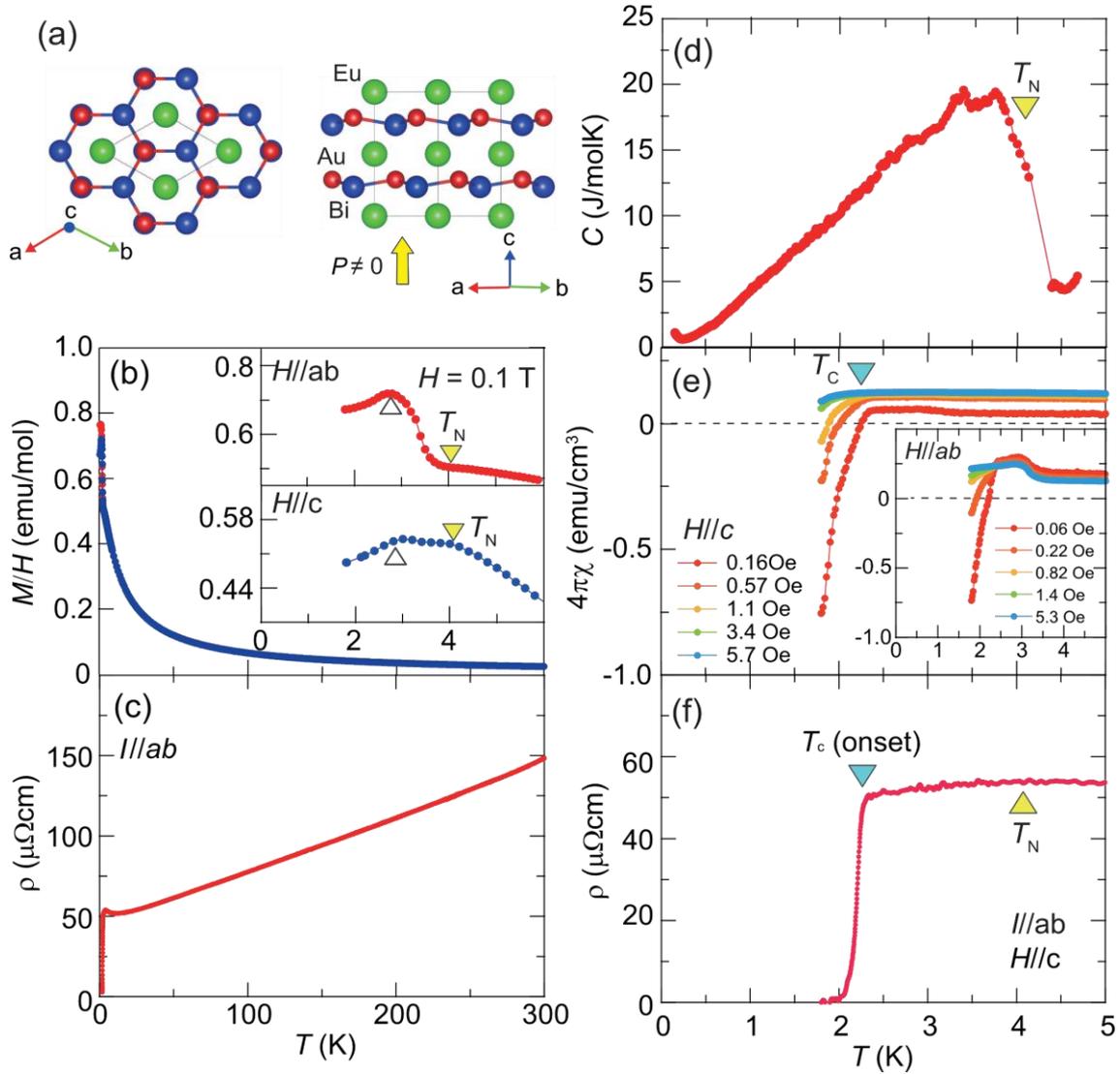

FIG. 1. (Color online) (a) Crystal structure of EuAuBi viewed along (left) and perpendicular to the *c* axis (right). (b) Temperature dependence of the magnetic susceptibility $M/H$ measured with $H = 0.1$ T for $H//c$ (blue) and $H//ab$ (red). Upper and lower insets show the magnetic susceptibility below 6 K with $H = 0.1$ T for $H//c$ and $H//ab$, respectively. (c) Temperature dependence of the in-plane electrical resistivity ρ. (d) Temperature dependence of specific heat capacity *C*. (e) The Meissner effect under the out-of-plane field ($H//c$) at various fields. Inset shows the Meissner effect under $H//ab$. (f) ρ-*T* curves below 5 K at selected magnetic fields.

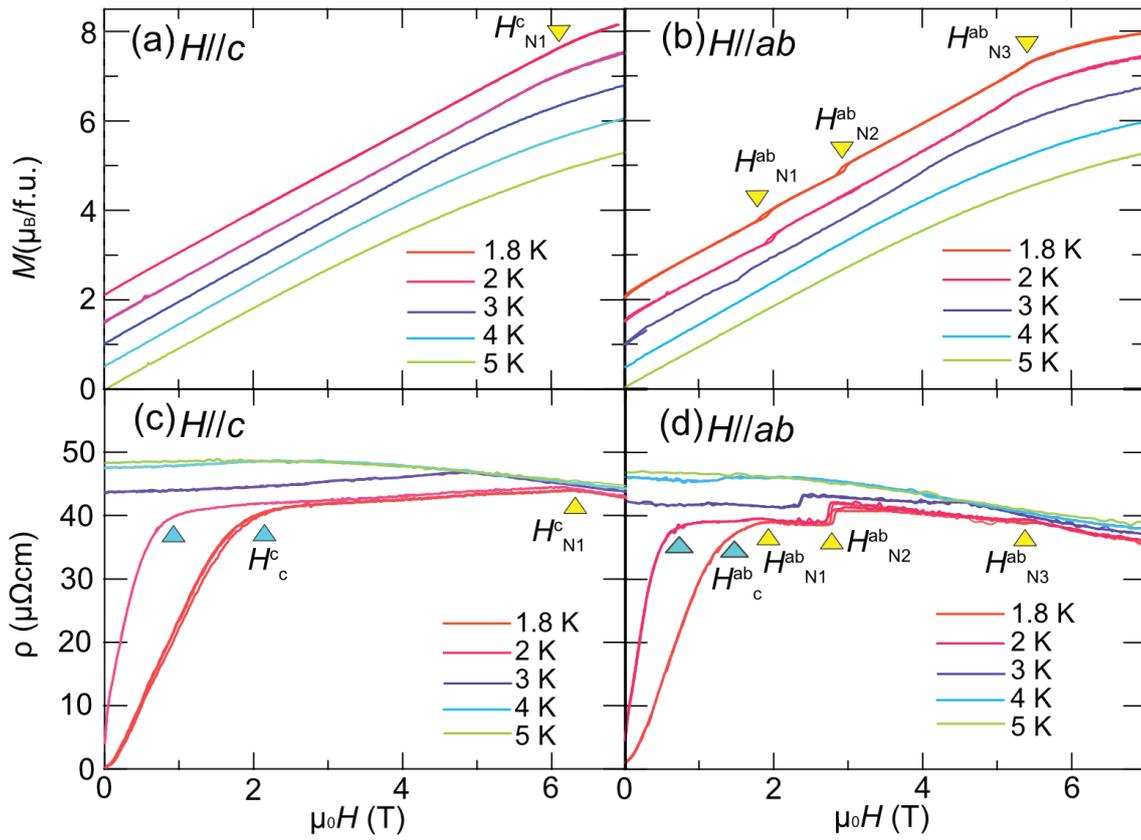

**FIG. 2.** (Color online) Magnetic field dependence of magnetization under (a) $H//c$ and (b) $H//ab$ below 5 K. Magnetic field dependence of in-plane resistivity under (c) $H//c$ and (d) $H//ab$ below 5 K.

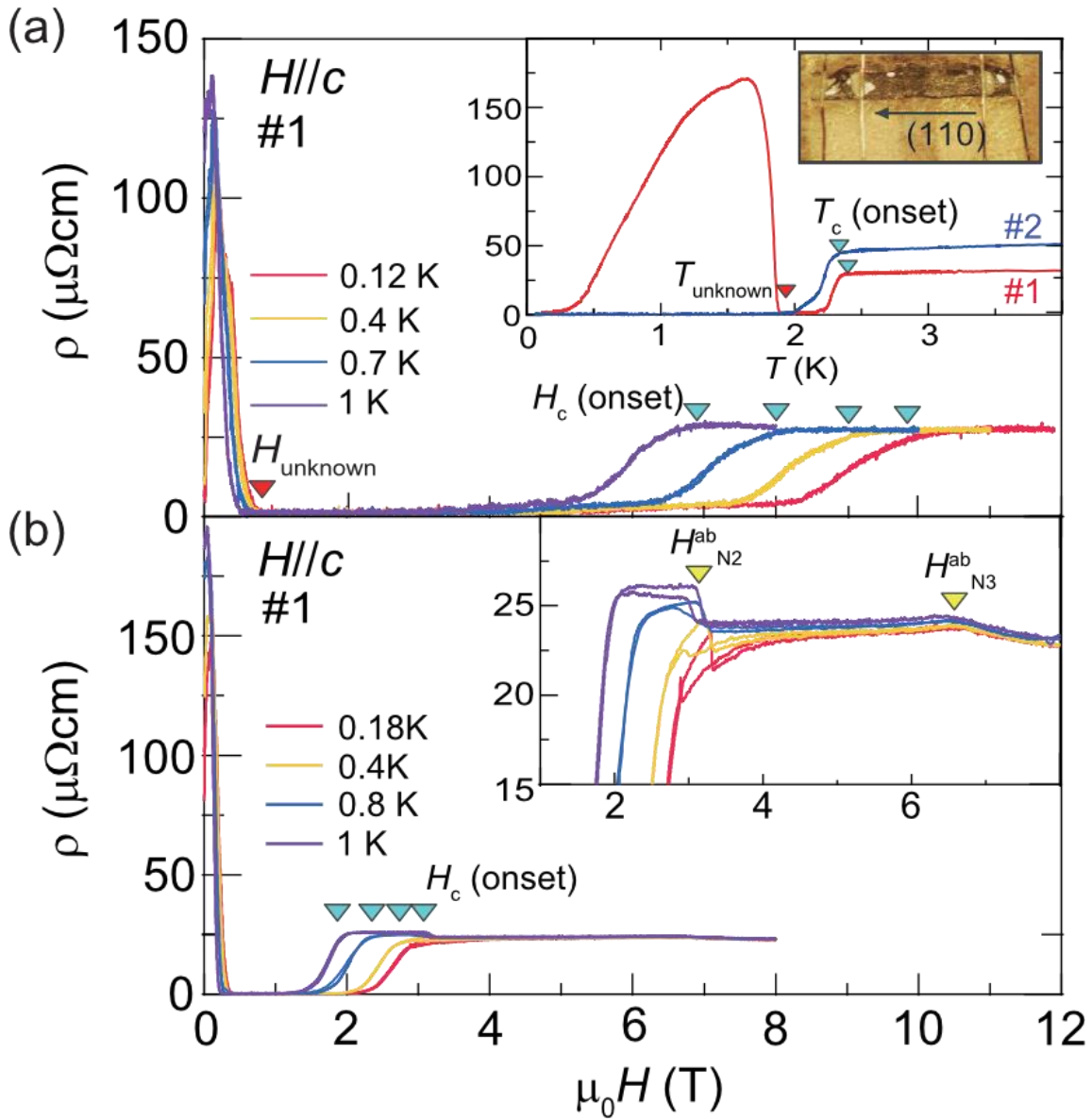

**FIG. 3.** (Color online) (a) Magnetic field dependence of in-plane resistivity of No. 1 under $H//c$ at selected temperatures. Inset shows temperature dependence of in-plane [$I//(110)$] resistivity from 4 to 0.1 K for two samples denoted as No. 1 and No. 2, and the photograph of a single crystal (No. 1) with lead wires. (b) Magnetic field dependence of in-plane resistivity of No. 1 under $H//ab$ at selected temperatures. Inset shows the magnified views of $\rho$-$H$ curves at high fields.

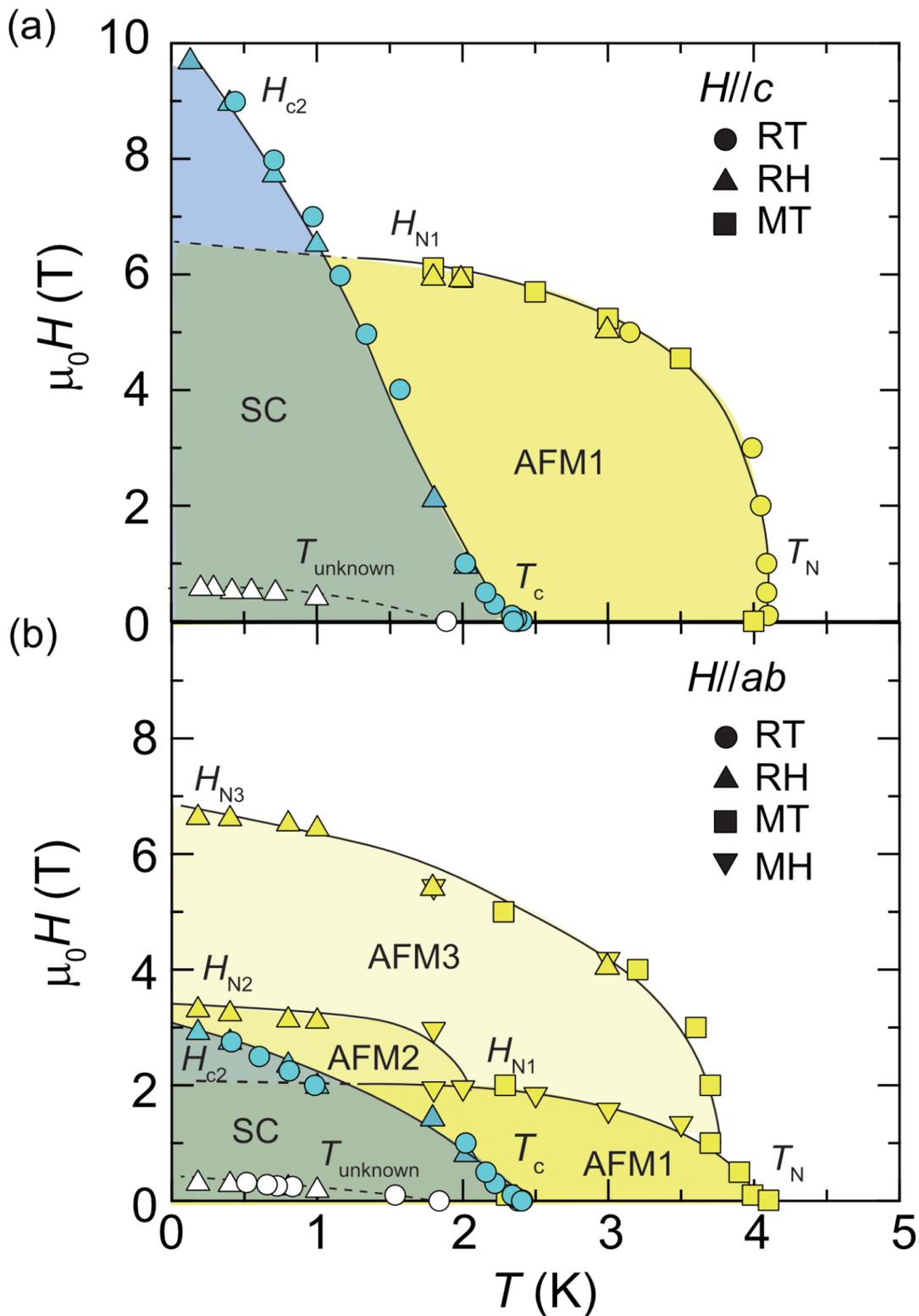

FIG. 4. (Color online) Magnetic and superconducting phase diagrams for (a) $H//c$ and (b) $H//ab$. Yellow and blue circles represent the magnetic transition temperature $T_N$ and the onset of superconducting (SC) transition temperature $T_c$, respectively. White circles denote the unknown anomalies observed in the resistivity of No. 1.

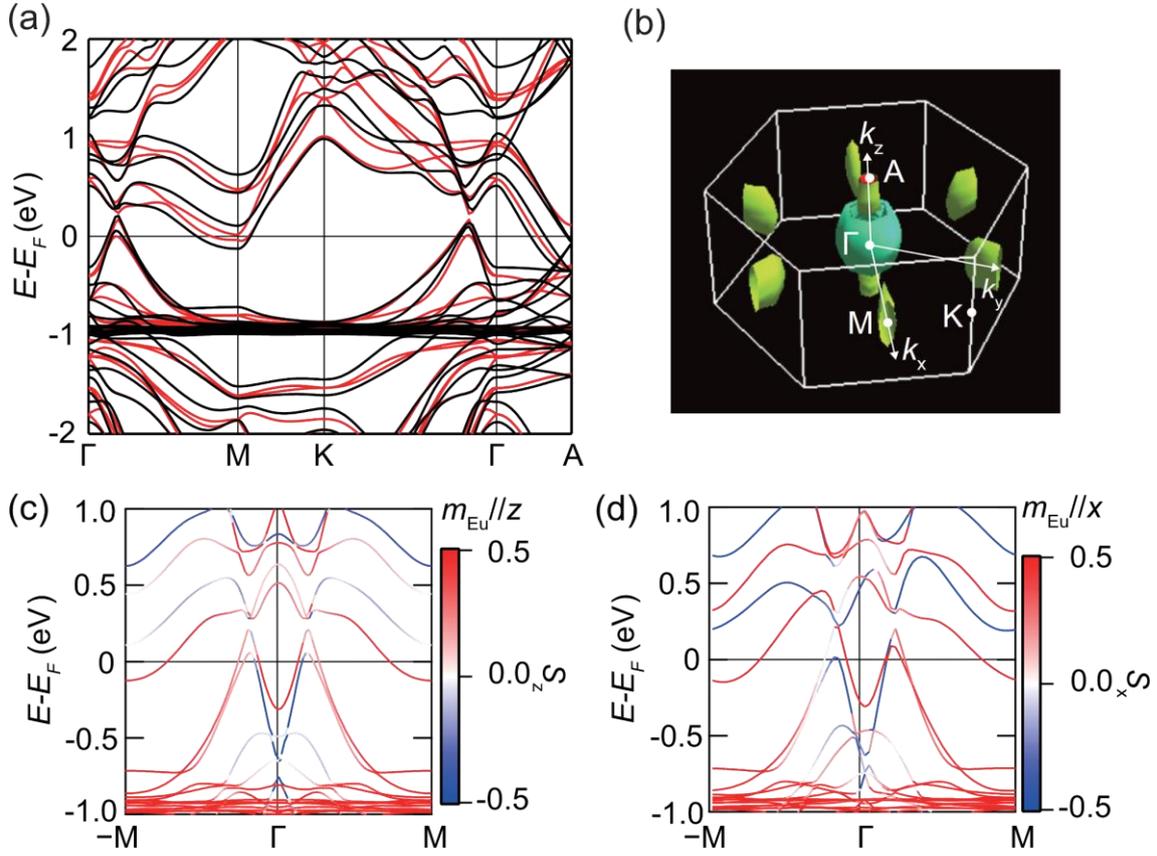

**FIG. 5.** (Color online) (a) The relativistic band structure of EuAuBi in FM (black) and AFM (red) phases with Eu moments parallel to the $c$ axis. (b) The Fermi surfaces in FM phase visualized by FermiSurfer package [46]. The calculated spin-polarization band dispersions in FM phase along -M–Γ–M for (c) $m_{Eu}//z$ and (d) $m_{Eu}//x$.